# Large-Core Optics for Simplified Short-Range FSO Links

**Florian Honz, and Bernhard Schrenk**
AIT Austrian Institute of Technology, Center for Digital Safety & Security / Security and Communication Technologies, 1210 Vienna, Austria.
Author e-mail address: florian.honz@ait.ac.at

**Abstract:** We evaluate large-core FSO links where excellent coupling pairs with bandwidth fading due to multi-mode propagation. The 10-Gb/s/λ limit for 105-µm double-clad fibers is mitigated by spectral launch tuning, restoring 84% of single-clad 25-Gb/s/λ capacity. © 2024 The Author(s)

## 1. Introduction

Free-space optical (FSO) communication is an attractive solution to preserve the bandwidth continuity in fiber-scarce environments. Applications traditionally foresee the bridging of fiber-grade transmission networks, with recent trials demonstrating capacities of more than 14 Tb/s [1] or 1.1 Tb/s/λ leveraging mode multiplexing [2]. FSO can further boost data rates in wireless access where sub-THz technology might face severe limitations [3]. In either case, simplicity is paramount: Interrupting or extending the fiber continuity comes at the cost of coupling light to and from an "ideal" waveguide-based transmission medium. This, apart from becoming vulnerable to weather conditions, requires complex assemblies for FSO beam shaping, acquisition, and tracking [4,5], which often renders FSO as an economically unattractive solution. As a way to offset deployment costs, we have recently proposed to augment traditional single-mode fiber (SMF) based FSO systems with a double-clad fiber (DCF) interface and demonstrated the robust relay of narrowband 250-MHz OFDM signals in an analogue mobile fronthaul link [6].

This work compares three large-core FSO configurations with respect to coupling efficiency and link bandwidth. We show that fading induced by mode filtering effects can be mitigated when altering the mode excitation through spectral tuning. We then demonstrate a passive short-range duobinary FSO link for 40 Gb/s out-door transmission.

## 2. FSO Link Leveraging Large-Diameter Core Coupling and Experimental Setup

Large-core fiber optics can improve the alignment tolerance and long-term stability of FSO links. However, multi-mode systems involve the coupling of light to higher-order modes, which inevitably leads to differential mode delay (DMD) as a bandwidth-limiting factor [7]. This impairment is enhanced by the lack of large-diameter high-speed photodetectors. Here, we will investigate broadband data transmission for three configurations of large-core FSO links (Fig. 1a). The first two employ an alignment-tolerant DCF layout with a single-mode launch core and a 105-µm inner cladding for light collection. These two configurations can be used in a bidirectional fashion and differ mainly by the 105 µm (preferred, ❶) or 200 µm (sub-optimal, ❷) core diameter of the multi-mode fiber (MMF) reception port. The third layout exclusively builds on 50-µm OM4 fiber (❸) at the receiving end, promising the highest link bandwidth, yet at the cost of a strictly unidirectional layout and reduced robustness for light coupling.

Figure 1b presents the experimental setup. The out-door FSO link is set up over a distance of 63 m and includes two fully-passive terminals with 2" optics interfacing the DCF couplers at the transmitting and receiving FSO units. At the transmitter side, we modulate four CWDM wavelengths from 1530 to 1590 nm with a Mach-Zehnder modulator (MZM) in a rate-adaptive scheme that adjusts the modulation parameters to the supported bandwidth of the three FSO channels under evaluation. The CWDM signals in the C- and L-band are then boosted by EDFAs and launched towards the single-mode port of the DCF coupler embedded within the roof-top installation.

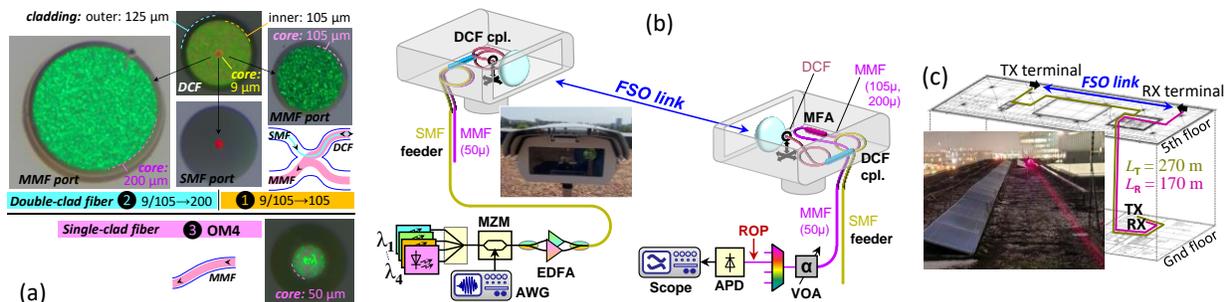

Fig. 1. (a) Large-core FSO fiber coupling. (b) Experimental setup. (c) Installation with roof-top FSO link and feeder fibers towards TX and RX.

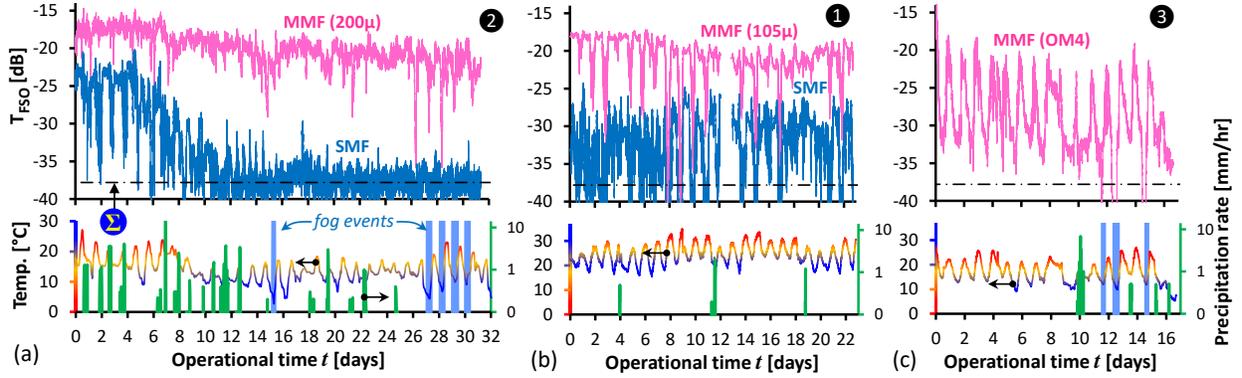

Fig. 2. FSO channel transmission for (a) DCF with 200-μm MMF and (b) 105-μm MMF used as receiving fiber. (c) Comparison to OM4 fiber.

At the receiver side, we employed mode-field adapters (MFA) for FSO configurations 1 and 2 before the coupled signal is fed to the lab premises using a 170-m OM4 MMF link. This feeder will play a key role in filtering spatial higher-order modes. For the third FSO configuration, light is directly coupled into the OM4 feeder link. At the end of this feeder, the signals are demultiplexed by multi-mode CWDM fiber-optics and detected by an avalanche photodetector (APD) with an active diameter of 33 μm. The transmission performance is then evaluated off-line as a function of the received optical power (ROP). Although the DCF-based configurations 1 and 2 support cross-talk robust bidirectional FSO communication, as proven earlier [6], our focus resides on a unidirectional link layout.

### 3. Characterization of Multi-Mode FSO: Coupling Efficiency, Alignment Tolerance and Link Bandwidth

Figure 2 reports the coupling stability for all FSO configurations, including on-site weather data. SMF coupling is rendered incompatible with the passive FSO terminals and would require active alignment. The strong temperature dependence is attributed to the thermal expansion of opto-mechanics. It is more pronounced for configuration 1 (Fig. 2b) due to the high peak-to-peak swing $\Delta T$ of 33.2°C inside the terminals. The coupling to OM4 fiber (configuration 3) shows the worst stability, despite a lower $\Delta T$ of 25.2°C. It is clearly impacted by day-to-night temperature cycles.

As the second performance criterion we investigated the frequency response, which is largely determined by the excitation of higher-order modes through either launch conditions or physical-channel discontinuities such as micro-bends, in combination with mode filtering effects. For this, we first coupled light between a single-mode launch core and a large-diameter reception core over a smaller free-space lab bench, having a collimated beam with a waist size of 2.8 mm. We then varied the beam offset and core angles relatively to each other, in absence of additional weather conditions. A 100-m span of OM4 fiber has been appended to the receiving fiber to emulate the feeder fiber as an additional mode filter before the APD receiver. Figure 3a presents the impact of a transmit beam offset when using OM4 and DCF (with 105-μm MMF) as receiving fibers. Even though the DCF fiber is more tolerant to an offset when compared to the OM4 fiber, its link response is already impacted without displacement. Towards that, we noticed a similar bandwidth support for DCF and OM4 without the mode filter introduced by the 100-m feeder fiber.

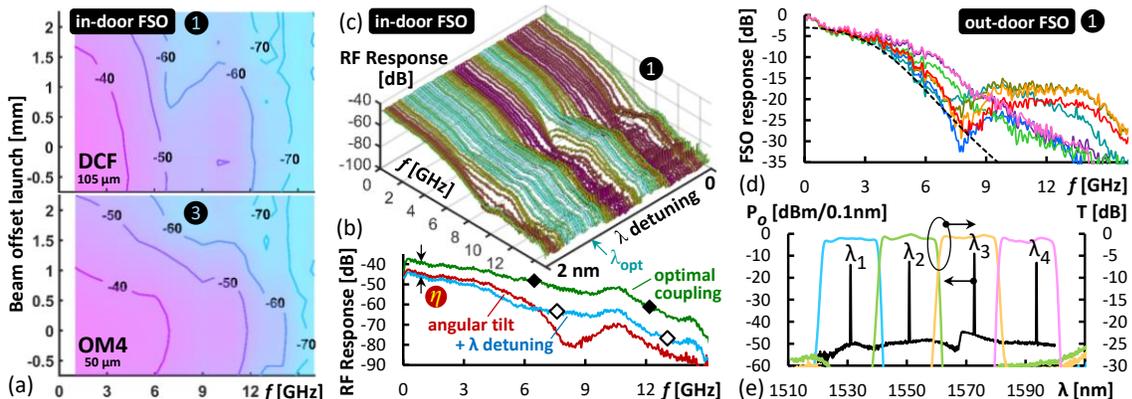

Fig. 3. In-door FSO bandwidth for (a) beam offset and (b) tilt, while (c) detuning λ. (d) Out-door FSO bandwidth. (e) Launched FSO channels.

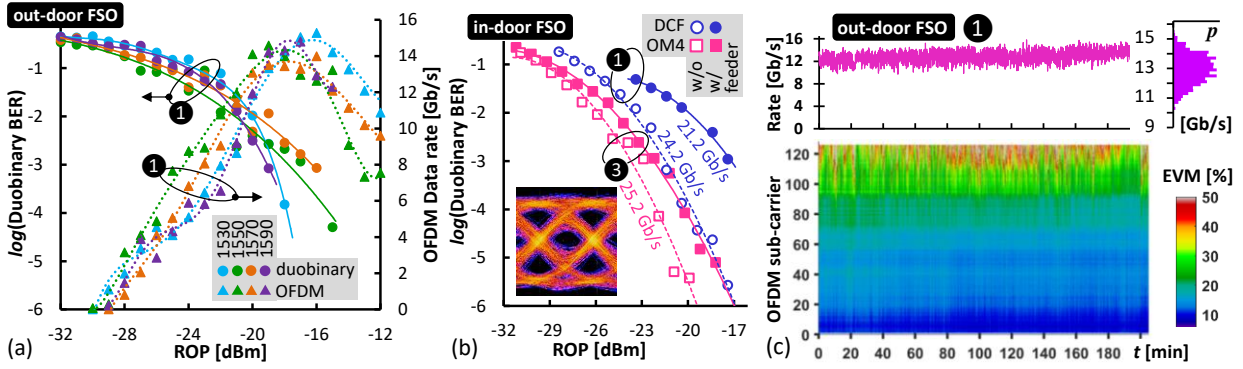

Fig. 4. Transmission performance for (a) out-door and (b) in-door FSO links. (c) Longer-term OFDM performance over out-door FSO link.

We see a stronger dependence for an angular tilt of the launch fiber core. As Fig. 3b shows, there is a spectral notch establishing when the optical axes are not perfectly aligned and higher-order modes are excited. To mitigate this penalty, we build on spectral tuning as a resource to shape the mode excitation by detuning the transmit laser wavelength by up to 2 nm for bandwidth-centric link optimization. Figure 3c reports the link responses for a step size of 3.8 GHz in detuning, indicating $\lambda_{opt}$ as the optimum: We can recover the frequency characteristics of the optimally-coupled FSO link (◆), as it is evidenced in Fig. 3b (◇). As can be expected, the reduced fiber-to-fiber coupling efficiency of 2.7 dB ($\eta$) cannot be compensated. This offset would have to be absorbed by the link margin.

We then acquired the end-to-end response including the out-door FSO link. Figure 3d reports captures at random time instances during windy weather. The frequency responses fall between the best- and worst-case conditions of the in-door FSO link, meaning that atmospheric propagation and vibrations result in the earlier investigated DMD-induced fading due to sub-optimal coupling to the large-core fiber. Since the variations now occur on a much faster time scale, the wavelength tuning scheme cannot be applied swiftly enough for the present experiment. For this reason, we have chosen a conservative link bandwidth (dashed line in Fig. 3d) to shape the modulation parameters for further link evaluation, according to a 4$^{th}$-order Bessel low-pass function with a -3dB bandwidth of 2.9 GHz.

## 4. Transmission Performance over Double- and Single-Clad FSO Links

We applied duobinary signaling at 10.7 Gb/s/λ over all four CWDM channels (Fig. 3e) as partial-response format matched to the FSO link response with DCF-based coupling to a 105-µm MMF receiving fiber (configuration 1). Figure 4a presents the BER performance (●). The reception sensitivity for the worst-case CWDM channel is -18.2 dBm at a BER of $3.8 \times 10^{-3}$. The spread in sensitivity among all CWDM channels of this 40-Gb/s link is 1.8 dB.

To make reference to a DMD-less case with fading-free channel, Fig. 4b shows the BER for the in-door FSO link with optimal coupling (Fig. 3b, ◆). For the same DCF with 105-µm MMF receiving fiber, we now achieve a sensitivity of -22.7 dBm at a higher-rate 21.2 Gb/s duobinary signal (●). FSO configuration 3 with the OM4 fiber (■) performs slightly better, which is attributed to a reduction in unfavorable mode coupling effects in the 100-m OM4 feeder before the APD. Here, a sensitivity of -23.8 dBm has been achieved for a 25.2 Gb/s duobinary signal.

We further evaluated the out-door FSO transmission for OFDM modulation at all four CWDM channels (Fig. 4a, ▲), using 128 adaptively modulated sub-carriers allocated over a spectrum of 5 GHz. We accomplished a peak post-FEC data rate of 14 Gb/s at a ROP of -16 dBm for the worst-case CWDM channel at 1570 nm, which marks the onset of APD saturation and degradation in error vector magnitude (EVM). Compared to duobinary signaling at a post-FEC data rate of 10 Gb/s, this point would be reached for a ROP of -20.7 dBm, meaning a sensitivity advantage of 2.5 dB for OFDM. However, since there is no deep feeding within the -10dB link bandwidth, the multi-carrier format doesn't offer a significant implementation advantage due to the conceptual simplicity of duobinary signaling.

Finally, Fig. 4c reports the OFDM performance at the 1550-nm channel over more than 3 hours. The EVM performance, which is closely related to the end-to-end link response, proves that a stable bit loading can be accomplished for a wind-shielded setup, yielding a steady post-FEC data rate of 12.8 Gb/s (3σ = 3.02 Gb/s).

## 5. Conclusion

We have investigated passive FSO links with favorable light coupling to large-core fibers, for which DMD-induced bandwidth limitations limit the FSO capacity to 10 Gb/s/λ for a 105-µm core fiber. However, bidirectional DCFs enable full-duplex setups, so that single-clad FSO layouts have to offset the cost of two parallel physical channels through an elevated performance. Given the suppression of bandwidth fading through spectral tuning, as shown for the DCF for in-door conditions at 84% of single-clad fiber capacity, we see the double-clad solution prevailing.

Acknowledgement: This work was supported by the ERC under the EU Horizon-2020 programme (grant agreement No 804769).